\documentclass[aps,pra,superscriptaddress,showpacs,showkeys,a4paper,10pt,notitlepage,twocolumn]{revtex4-2}
\usepackage[utf8]{inputenc}
\usepackage[english]{babel}
\usepackage{amsmath}
\usepackage{amssymb}
\usepackage{amsfonts}
\usepackage{graphicx}

\usepackage{braket}
\usepackage{float}
\usepackage{natbib}
\usepackage{cancel}
\usepackage{upgreek}
\usepackage{mathtools}
\usepackage[T2A]{fontenc}

\usepackage{xcolor}
\definecolor{link_blue}{RGB}{52,46,157}

\usepackage{longtable}
\usepackage[colorlinks,citecolor=link_blue,linkcolor=link_blue,urlcolor=link_blue]{hyperref}
\usepackage[tiny,center,uppercase]{titlesec}

\setcitestyle{super}
\usepackage{tikz,xcolor}
\definecolor{lime}{HTML}{A6CE39}
\DeclareRobustCommand{\orcidicon}{%
 \begin{tikzpicture}
    \draw[lime, fill=lime] (0,0)
    circle [radius=0.16]
    node[white] {{\fontfamily{qag}\selectfont \tiny ID}};
    \draw[white, fill=white] (-0.0625,0.095)
    circle [radius=0.007];
    \end{tikzpicture}
    \hspace{-2mm}
}

\begin{document}

\title{Ground state of the gauge invariant Dicke model: condensation of the photons in non-classical states }

\newcommand{\orcidIDF}
{\href{https://orcid.org/0000-0003-0476-8634}{\orcidicon}}

\newcommand{\orcidODS}
{\href{https://orcid.org/0000-0002-2875-0140}{\orcidicon}}

\newcommand{\orcidSan}
{\href{https://orcid.org/0000-0003-2919-5414}{\orcidicon}}

\newcommand{\orcidAU}
{\href{https://orcid.org/0000-0002-7122-0954}{\orcidicon}}

\newcommand{\orcidLeonau}
{\href{https://orcid.org/0000-0002-4830-6856}{\orcidicon}}

\author{N.\ Q.\ San\orcidSan}
\email[Corresponding author: ]{nguyenquangsan@hueuni.edu.vn}
\affiliation{School of Engineering and Technology - Hue University, Hue, Vietnam}

\author{O.\ D.\ Skoromnik\orcidODS}
\affiliation{Currently without university affiliation}

\author{A.\ P.\ Ulyanenkov\orcidAU}
\affiliation{Atomicus GmbH Amalienbadstr. 41C, 76227 Karlsruhe, Germany}

\author{A.\ U.\ Leonau\orcidLeonau}
\affiliation{Deutsches Elektronen-Synchrotron DESY, Hamburg 22607, Germany
}

\author{I.\ D.\ Feranchuk\orcidIDF}
\affiliation{Atomicus GmbH Amalienbadstr. 41C, 76227 Karlsruhe, Germany}

\begin{abstract}
  We investigate the ground state of two physically motivated
  modifications of the Dicke model. The first modification corresponds
  to particles whose phase space contains only two states, for
  example, particles with spin 1/2 or artificially created qubits. The
  second modification describes two-level systems that arise as a
  result of truncating the full Hilbert space of atoms to two levels
  that are in resonance with the electromagnetic field and are
  described by the gauge-invariant Dicke model. We demonstrate that
  the behavior of these systems is qualitatively distinct in both
  cases. In particular, in the first scenario, a phase transition into
  the state with a non-zero amplitude of the classical field is
  possible, while in the second case, the so-called order parameter
  $ \eta = \braket{\hat{a}}$ of the field's phase transition into a coherent state
  with photon condensation is zero. At the same time, the average number of
  photons $\bar{n} = \braket{\hat{a}^\dagger \hat{a}} \neq 0$, and the collective excitation in
  the system manifests a non-classical "squeezed" state of the
  field. We analyze the observable characteristics of both systems in
  a wide range of variation of their parameters.
\end{abstract}

\pacs{}
\keywords{}
\maketitle

\section{Introduction}
\label{sec:introduction}

The Dicke Model (DM) \cite{PhysRev.93.99} describes the interaction
between light and matter and was introduced to describe coherent
processes in a dense gas induced by the electromagnetic field of a
resonator \cite{GROSS1982301}. An important peculiarity of this model
is the occurrence of a phase transition into the superradiant state,
corresponding to the excitation of a macroscopic number of photons in
the resonator. This means that the mean field amplitude $\braket{\hat{a}} \neq 0$
\cite{PhysRevB.100.121109}.

Another property of the DM is the existence of a phase
transition in a state of thermodynamic equilibrium of this system, which has drawn considerable attention starting from
works \cite{HEPP1973360,PhysRevA.7.831}. It was shown that in the zero
temperature limit, when the system is in its ground state, the phase
transition corresponds to the emergence of the superradiant phase,
which is associated with the formation of a photon condensate
\cite{PhysRevLett.92.073602,PhysRevLett.94.163601}.

However, it was demonstrated\cite{PhysRevLett.35.432,PhysRevA.19.301}
that when the DM is used to describe the real atoms, i.e. atoms whose
spectrum contains many levels, it is necessary to account for the
diamagnetic contribution in the Hamiltonian from the interaction
between the atoms and the field. This leads to the emergence of a quadratic term in the
model Hamiltonian involving field operators and prohibits the
thermodynamic phase transition in this system, as proven by the no-go theorem \cite{PhysRevB.100.121109,andolina22}. Similar statements have also been established for a phase
transition in the ground state of the system leading to the formation of a
photon condensate. It is important to stress that proofs of the corresponding no-go theorems are  based on
transforming the Hamiltonian of the DM into a gauge-invariant
form, with the subsequent truncation of the dimensity of the Hilbert space to only two
levels \cite{di_stefano_resolution_2019,PhysRevLett.132.073602}.

At the same time, when using the DM to describe the real physical two-level systems -
qubits - a phase transition is allowed
\cite{PhysRevLett.92.073602,PhysRevLett.94.163601}. In particular, for systems with large finite numbers of qubits ($N \gg 1$) this problem has been
investigated numerically, and characteristics of the
phase transition have been identified \cite{PhysRevA.78.051801}. It corresponds to the
appearance of a classical field component in the system, characterized
by a non-zero order parameter $\eta = \braket{\hat{a}} \neq 0$. Recently, such a
state in the system with $N=1$ has been observed experimentally
\cite{PhysRevLett.131.113601}.

These results lead to the ambiguity of using the DM, i.e. whether it is exploited to describe simplified atomic systems with many levels being truncated to only two of them, or real two-state
systems as qubits, which naturally contain only two levels.

The aim of the present work is to investigate numerically the ground state properties of the gauge-invariant Dicke model (GIDM) \cite{di_stefano_resolution_2019} accounting for the multi-level energy spectrum of atoms, and compare them with the corresponding results emerging from the DM. The paper is organized as follows: in Sec. \ref{sec:hamiltonians-dm-gidm} we introduce the Hamiltonian of the DM and GIDM and describe the algorithm for numerical calculation of the ground state energies and wave functions of these systems. In Sec. \ref{sec:observ-char} other characteristics of the ground states are calculated and the possibility of phase transition is analyzed. In Sec. \ref{sec:conclusion} we discuss the obtained results.


\section{Hamiltonians of DM and GIDM}
\label{sec:hamiltonians-dm-gidm}

Let's consider two models used to describe the resonant interaction of
radiation with atoms in a resonator. The first model is the standard
DM with the Hamiltonian corresponding to a system
consisting of $N$ qubits interacting with a single-mode quantum field \cite{PhysRevA.78.051801}:
\begin{eqnarray}
\label{1}
\hat{H}_D = \hat{a}^\dagger \hat{a} + \Delta \hat{J}_z + \frac{2\lambda}{\sqrt{N}}(\hat{a} + \hat{a}^\dagger) \hat{J}_x + \lambda^2,
\end{eqnarray}

\noindent where the field frequency $\omega$ is considered as the energy unit; $\Delta$ is the transition frequency between the qubit levels; $\hat{a}$, $\hat{a}^\dagger$ are the annihilation and creation operators of the quantum field with the frequency $\omega = 1$; $\hat{J}_z$ and $\hat{J}_x$ denote the projections of the total
angular momentum $J = \frac{N}{2}$; $\lambda$ is the coupling constant between the
qubit and the field. Here we use the system of units $\hbar = c = 1$.

The coupling constant can be defined as follows:
\begin{eqnarray}
\lambda = d \sqrt{2\pi \rho},
\end{eqnarray}	

\noindent where $d$ is the dipole moment of the qubit, and $\rho$ is the density
of qubits in the resonator.

The second model is the GIDM  corresponding to a system
consisting of $N$ atoms, whose energy spectrum is truncated to only two levels, interacting with a single-mode quantum field \cite{di_stefano_resolution_2019}:
\begin{eqnarray}
\label{2b}
\hat{H}_C = \hat{a}^\dagger \hat{a} + \Delta \biggl\{ \hat{J}_z \cosh \left[ 2f (\hat{a} - \hat{a}^\dagger) \right]  \nonumber\\
+  i \hat{J}_x \sinh \left[ 2f (\hat{a} - \hat{a}^\dagger) \right] \biggr\}  .
\end{eqnarray}

We note that canonical transformations are made to match the
parameters in the operator from \cite{di_stefano_resolution_2019} with
the form of Hamiltonian $\hat{H}_D$:
\begin{eqnarray}
\hat{a} \rightarrow  i \hat{a}; \quad \hat{a}^\dagger \rightarrow -i \hat{a}^\dagger; \quad \hat{J}_y \rightarrow \hat{J}_x,
\end{eqnarray}

\noindent with the coupling constant $f$ having another normalization
\begin{eqnarray}
\label{2a}
f = \frac{\lambda}{\sqrt{N}}.
\end{eqnarray}

In addition,  the term $\lambda^2$ is introduced in (\ref{1}) to
provide the same energy reference level as in the operator (\ref{2b})
\cite{di_stefano_resolution_2019}.

In order to solve the Schrödinger equation for (\ref{1}) and (\ref{2b}) numerically, we use the following expansion for the eigenvectors of the system's states:
\begin{eqnarray}
\label{3}
\ket{\Psi} = \sum_{n=0}^{N_{tr}} \sum_{M=-N/2}^{N/2}C_{nM} \ket{n} \chi_{J,M},
\end{eqnarray}
where $\ket{n}$ is the Fock state of the field, $N_{tr}$ the upper value of the photon state taken into account; $\chi_{J,M}$ eigenfunctions of the angular momentum operator:
\begin{eqnarray}
\hat{J}^2 \chi_{J,M} = J(J+1)\chi_{J,M}; \quad \hat{J}_z \chi_{J,M} = M\chi_{J,M}.
\end{eqnarray}

Using $\ket{n}$ and $\chi_{J,M}$ as the basis vectors, we derive the following expressions for the matrix elements of the Hamiltonian (\ref{1}):
\begin{eqnarray}
\label{4}
(H_D)_{kn}^{MM'} = \left(n \delta_{nk} + \Delta M + \lambda^2 \right) \delta_{MM'}  +  \nonumber \\ \frac{\lambda}{\sqrt{N}}
\left(\sqrt{n}\delta_{n-1,k} +\sqrt{n+1}\delta_{n+1,k}\right)\times \nonumber\\
 \sqrt{(J+M)(J-M+1)}\left(\delta_{M-1,M'} +\delta_{M+1,M'}\right),
\end{eqnarray}
and the Hamiltonian (\ref{2b})\cite{leonau2022eigenstates}:
\begin{align}
\label{qrm-matr}
H_{kn}^{MM'} = n \delta_{kn} \delta_{MM'} + \frac{\Delta}{2} S_{kn} \biggl[ \frac{(-1)^n + (-1)^k}{2} M\delta_{MM'} \nonumber \\
	 - i \frac{(-1)^n - (-1)^k}{2} \sqrt{(J+M)(J-M+1)}\times \nonumber\\(\delta_{M-1,M'} +\delta_{M+1,M'}) \Bigr],\nonumber \\
	S_{kn}(f) = (-1)^n  \sqrt{\frac{n!}{k!}} (2f)^{k-n}L_n^{k-n}(4f^2) e^{-2f^2}; \nonumber \\
	 k \geq n; \ S_{kn}= S_{nk},
\end{align}
where $L_n^k (x)$ are the generalized Laguerre polynomials.

\section{Observable characteristics}
\label{sec:observ-char}

\begin{figure*}[tb!]
	\includegraphics[width=0.95\textwidth]{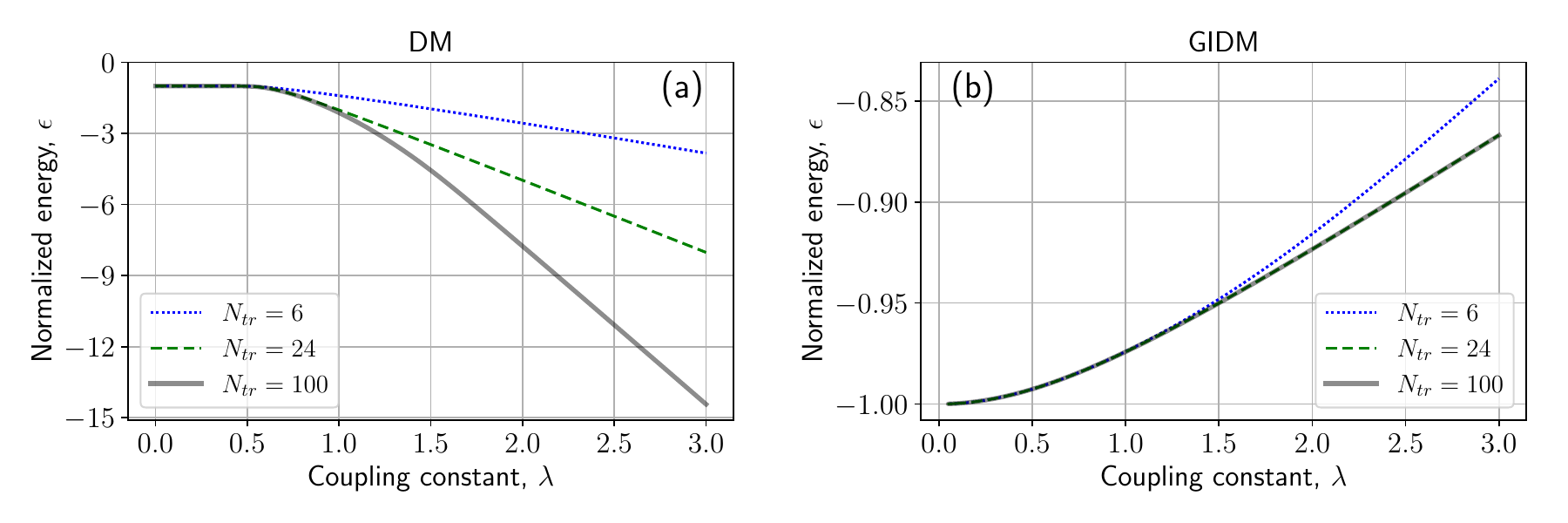}
	\caption{(Color online) Normalized ground state energy $\epsilon$  of the (a) DM and (b) GIDM with $N=32$ atoms/qubits and $\Delta = 1$ as a function of the coupling constant $\lambda$ for the upper photon truncation limits: $N_{tr}=6$ (dotted line), $N_{tr}=24$ (dashed line), $N_{tr}=100$ (solid line). Results from Fig.~\ref{fig:1}a coincide with the results of Fig.~1 in \cite{PhysRevA.78.051801} .}
	\label{fig:1}
\end{figure*}
\begin{figure*}[tb!]
	\includegraphics[width=0.95\linewidth]{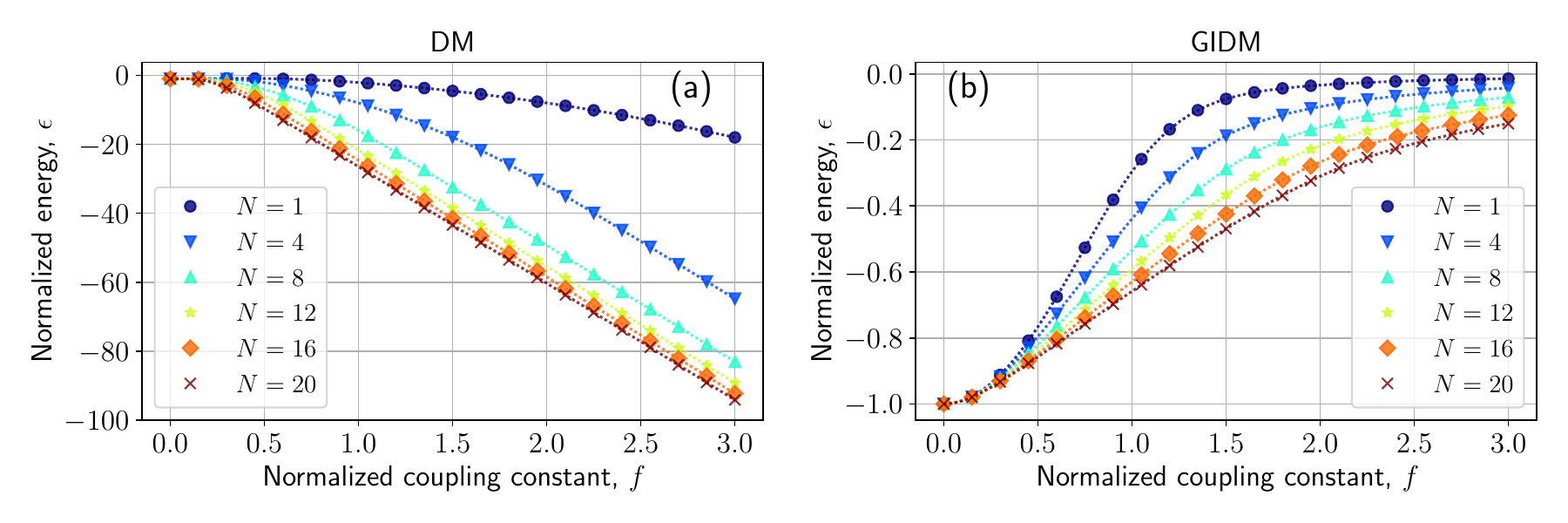}
	\caption{(Color online) Normalized ground state energy $\epsilon$ of the (a) DM and (b) GIDM as a function of the normalized coupling constant $f$ with different numbers of atoms/qubits $N$ and $\Delta=1$.}
	\label{fig:2}
\end{figure*}

\begin{figure}[tb!]
	\includegraphics[width=0.90\linewidth]{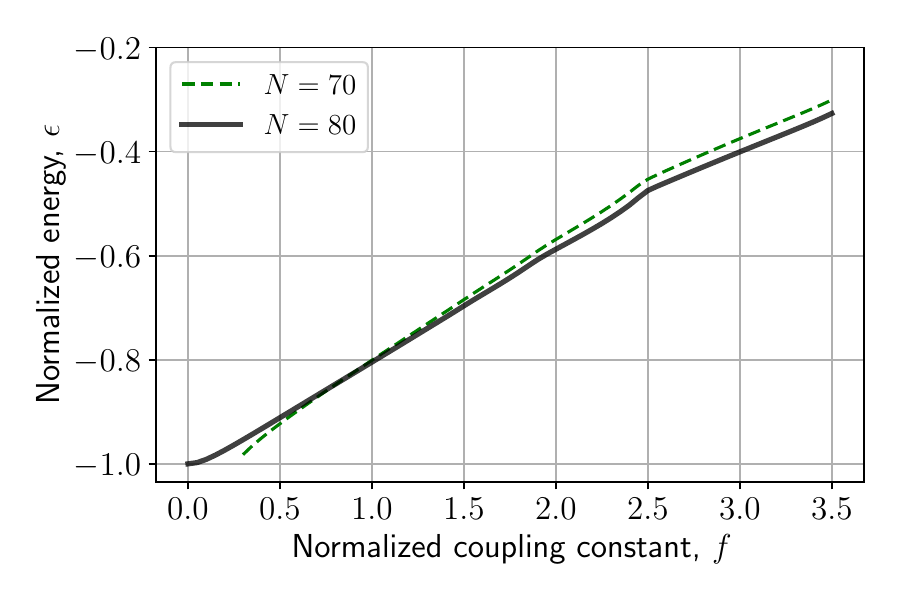}
	\caption{(Color online) Accuracy  of the functional equation (\ref{6}): normalized ground state energy $\epsilon$ of the GIDM as a function of the normalized coupling constant $f$ and $N=70$ (dotted line), $N=80$ (solid line) and $\Delta=1$.}
	\label{fig:3}
\end{figure}
\begin{figure}[tb!]
	\includegraphics[width=0.90\linewidth]{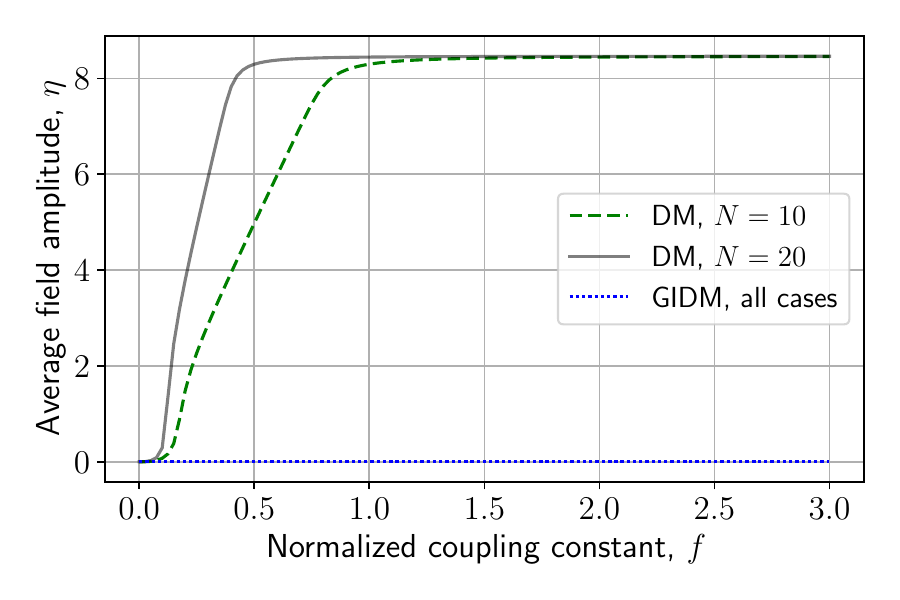}
	\caption{(Color online) Average amplitude of the field $\eta$ as a function of the normalized coupling constant $f$ for the ground state of the (a) DM and (b) GIDM and $\Delta=1$.}
	\label{fig:4}
\end{figure}

In Fig.~\ref{fig:1} we show the results of calculating the normalized values of the  ground state energies
\begin{equation}
	\epsilon (N,\lambda) = \frac{2 E_0(N,\lambda)}{N}
	\label{eq:ngre}
\end{equation}
\noindent  by numerical diagonalization of matrices (\ref{4}) and
(\ref{qrm-matr}). It is remarkable that results from Fig.~\ref{fig:1}a coincide with the results of Fig.~1 in \cite{PhysRevA.78.051801}. Let us also stress that for the GIDM sufficiently fast convergence rate with increasing the photon threshold value $N_{tr}$ is observed as shown in Fig.~\ref{fig:1}b.

Fig.~\ref{fig:2} presents the results of calculating the normalized
ground state energy of the system
as a
function of the coupling constant and the number of qubits/atoms in the
resonator for both models. For the DM (Fig.~\ref{fig:2}a) the decrease in energy with increasing $f$ is due to the increase in the binding energy of  the qubits as a result of the radiation-simulated attractive potential discussed in our work \cite{PhysRevA.102.043702}. Fig.~\ref{fig:2}b shows that for the GIDM, with increasing the coupling constant (atomic density), part of the atoms transits to the excited state, which is consistent with the results of our work \cite{leonau2022eigenstates}.

We note that for the GIDM and  $N \gg 1$ the main contribution to
the energy is determined by the following term in the Hamiltonian
(\ref{2b}):
\begin{eqnarray}
H_1 \sim N e^{-2f^2},
\end{eqnarray}

\noindent which allows one to derive approximate analytical expression for the
energy for this particular case:
\begin{eqnarray}
\label{6}
 \epsilon (N,f)\approx \epsilon (N_1,f_1) + O \left(\frac{1}{N}\right); \nonumber\\ f_1 = \sqrt{f^2 - \frac{1}{2} \ln \frac{N}{N_1}}, \
 N>N_1; \ f^2 > \frac{1}{2} \ln \frac{N}{N_1}.
\end{eqnarray}

Fig.~\ref{fig:3} demonstrates sufficiently high accuracy of the
relation (\ref{6}), allowing one to determine the ground state
energy for any value of $N \gg 1$ by analytical continuation.

\begin{figure*}[tb!]
	\includegraphics[width=0.95\linewidth]{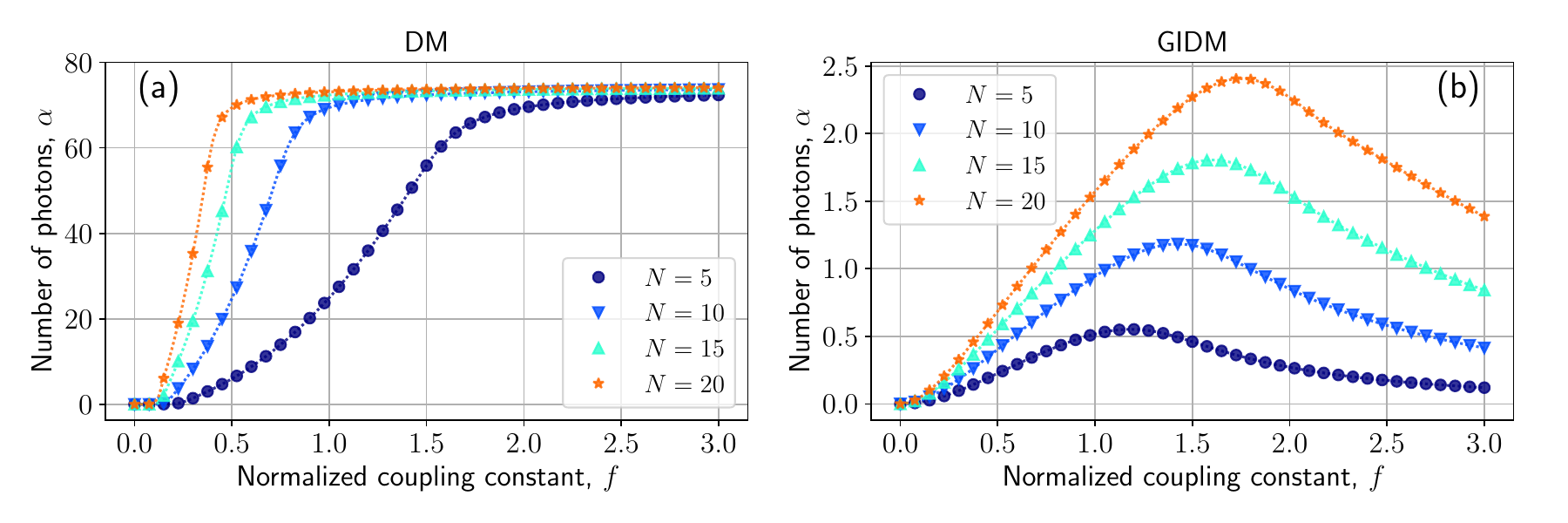}
	\caption{(Color online) Average number of photos $\alpha$ as a function of the normalized coupling constant $f$ for the (a) DM and (b) GIDM and $\Delta=1$.}
	\label{fig:5}
\end{figure*}
\begin{figure*}[tb!]
	\includegraphics[width=.98\linewidth]{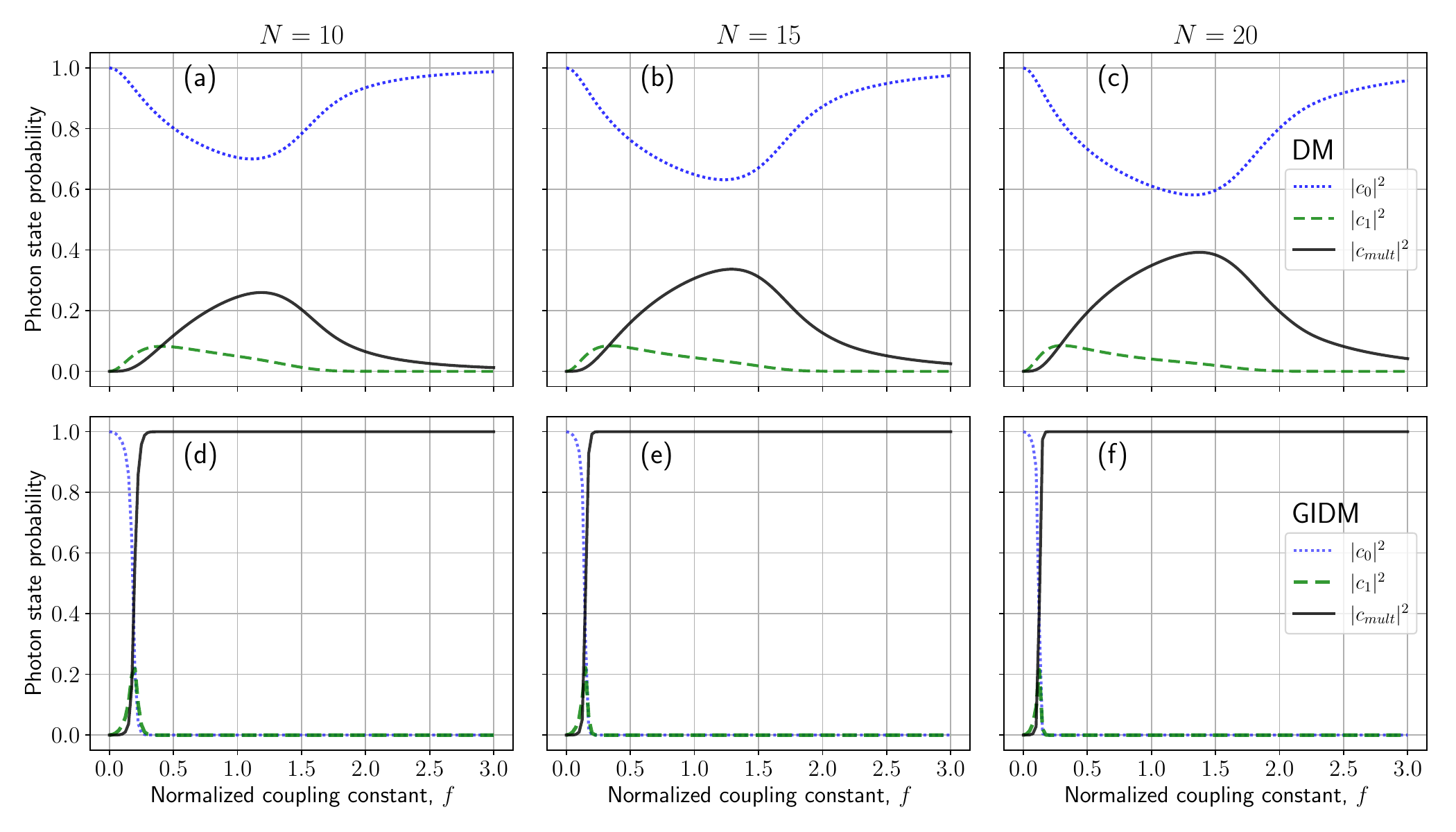}
	\caption{(Color online) Photon state probability $|c_n|^2$ as a function of the normalized coupling constant $f$ for the (a)--(c) DM and (d)--(f) GIDM and different number of atoms/qubits:
	(a), (d) $N=10$; (b), (e) $N=15$; (c), (f): $N=20$. The case $\Delta=1$ is shown.}
	\label{fig:6}
\end{figure*}

Let us investigate some observable characteristics of the system
being in the ground state. We start with the average amplitude
of the field:
\begin{eqnarray}
\label{7}
\eta =\braket{\Psi|\hat{a}|\Psi} =  \sum_n\sum_M \sqrt{n}C^*_{n-1,M} C_{nM}.
\end{eqnarray}

\begin{figure*}[tb]
	\includegraphics[width=0.98\linewidth]{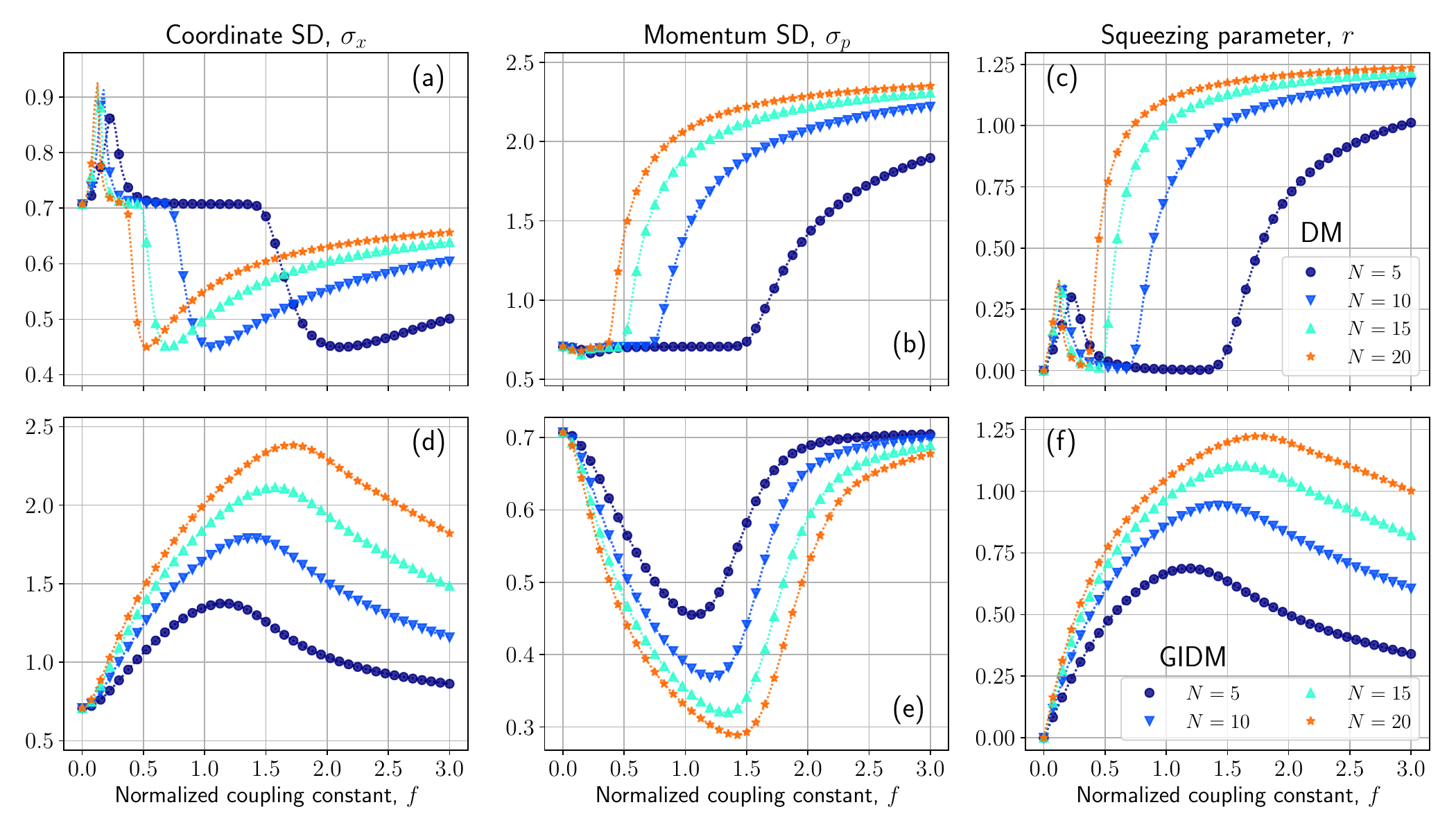}
	\caption{(Color online) Standard deviation of the coordinate $\sigma_x$; standard deviation of the momentum $\sigma_p$; and squeezing parameter $r$  as a function of the normalized coupling constant $f$ with different numbers of atoms/qubits $N$ and $\Delta=1$ for the (a)--(c) GM and (d)--(f) GIDM.}
	\label{fig:7}
\end{figure*}
\begin{figure*}[tb]
	\includegraphics[width=0.44\linewidth]{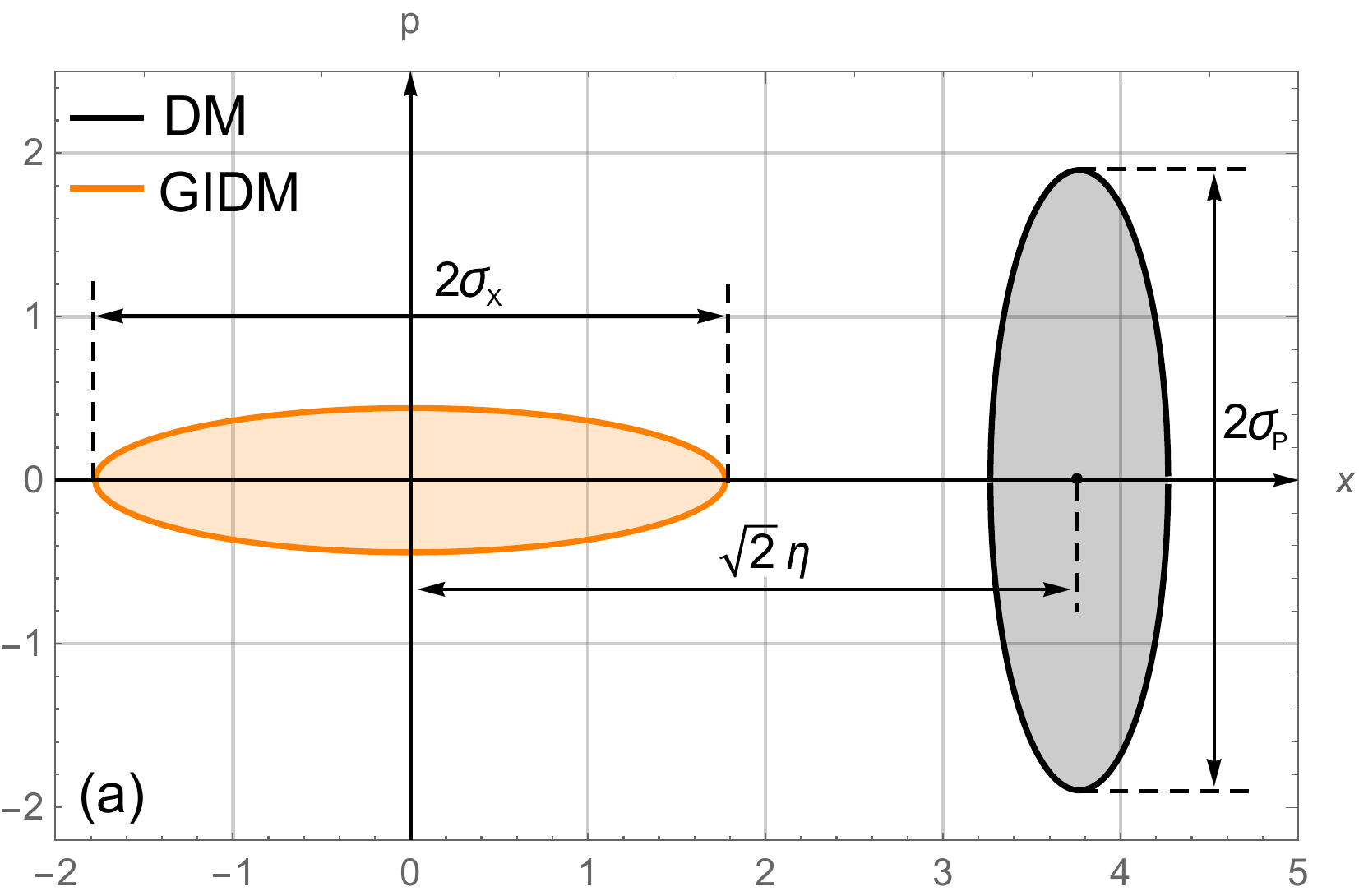}
	\quad\quad\quad
	\includegraphics[width=0.44\linewidth]{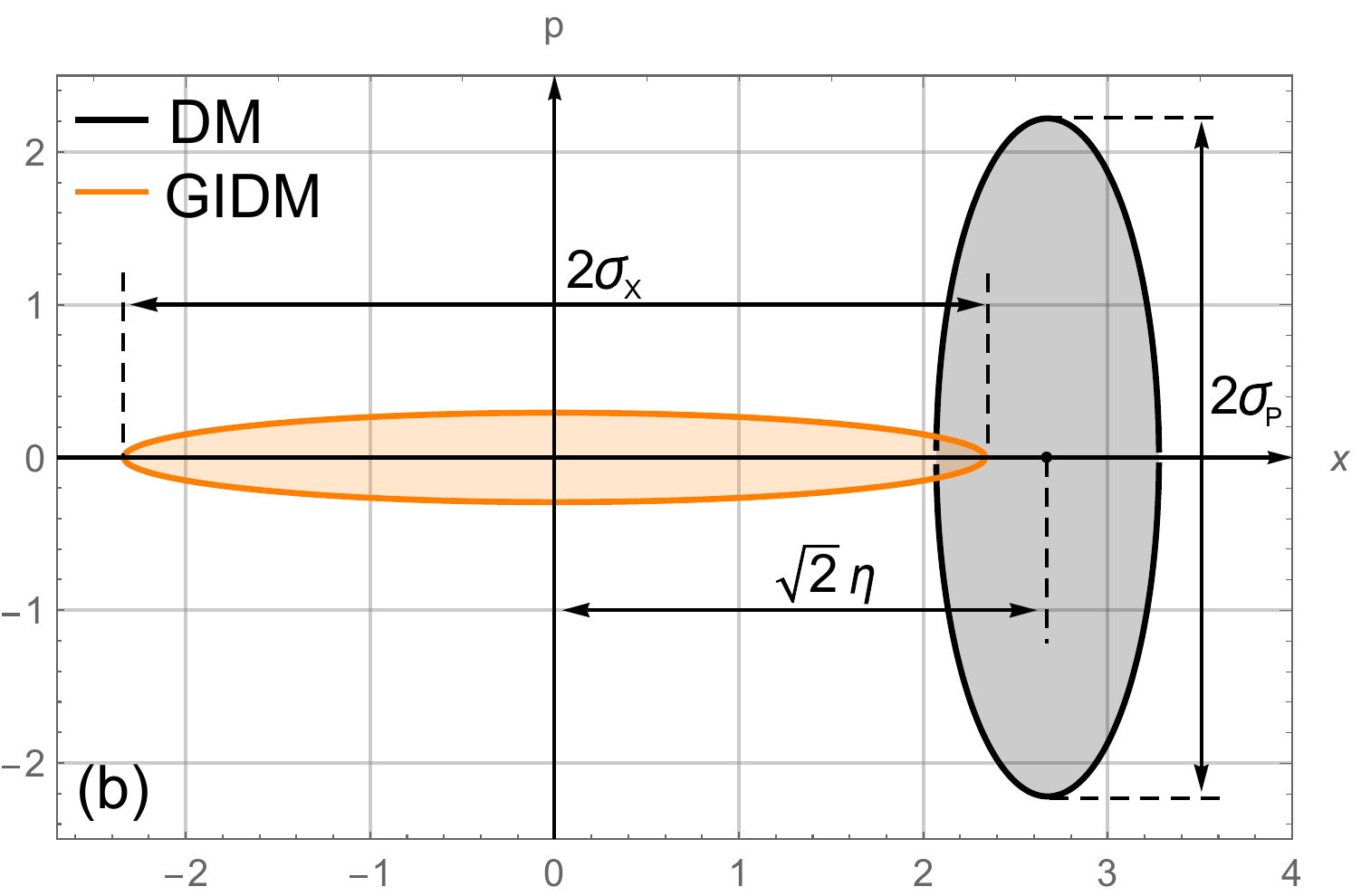}
	\caption{(Color online) Phase space of the photon condensate for the DM and the GIDM with (a) $N=10$, (b) $N=20$ atoms/qubits and $\Delta=1$.}
	\label{fig:phasespace}
\end{figure*}

\begin{figure*}[tb]
	\includegraphics[width=0.95\linewidth]{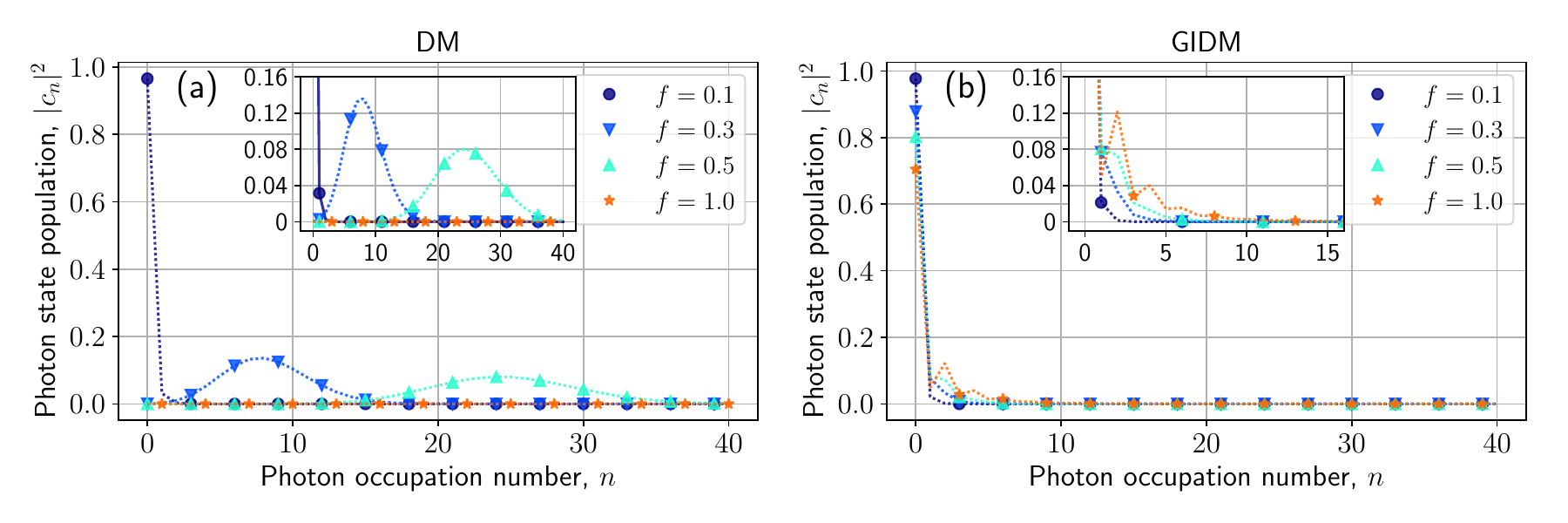}
	\caption{(Color online) Populations of the photon states $|c_n|^2$ at different normalized coupling constants for the (a) DM and (b) GIDM and $N=10$, $\Delta=1$. Both insets show the scaled view of the main figures.}
	\label{fig:8}
\end{figure*}
\begin{figure*}[tb]
	\includegraphics[width=0.95\linewidth]{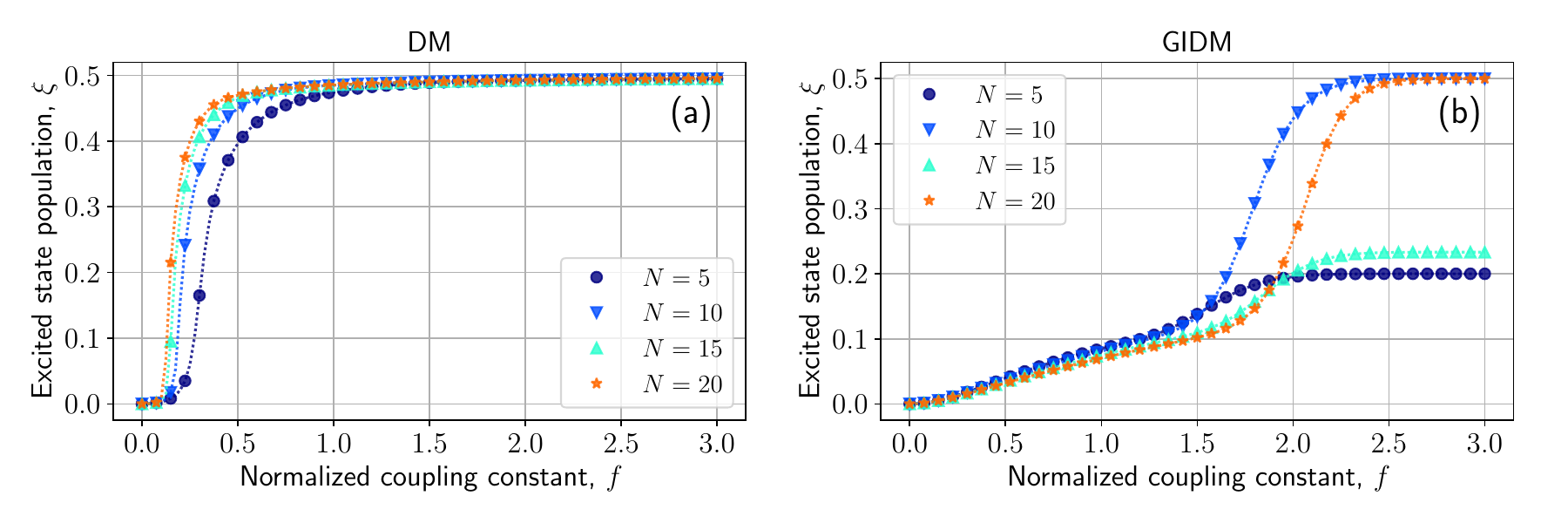}
	\caption{(Color online) Population of the excited state $\xi$ of atoms/qubits  as a function of the normalized coupling constant $f$  for the (a) DM and (b) GIDM with different numbers of atoms/qubits $N$ and $\Delta=1$.}
	\label{fig:9}
\end{figure*}
\begin{figure*}[tb]
	\includegraphics[width=0.95\linewidth]{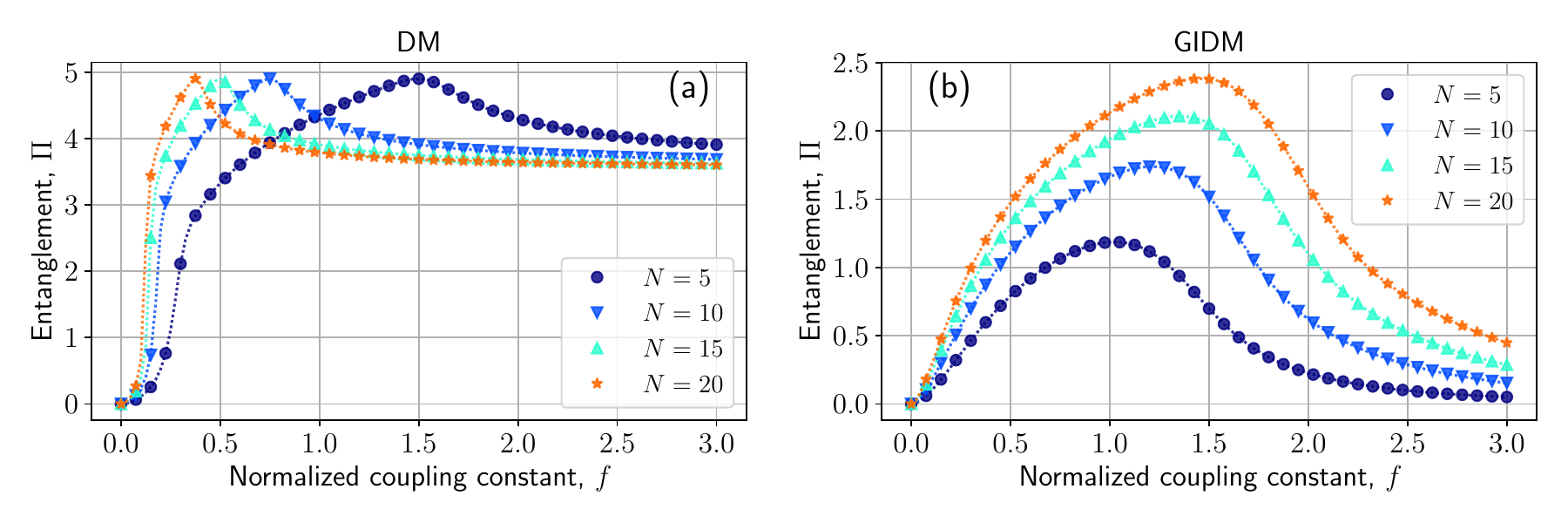}
	\caption{(Color online) Entanglement $\Pi$ of atomic and field states as a function of the normalized coupling constant $f$  for the (a) DM and (b) GIDM with different numbers of atoms/qubits $N$ and $\Delta=1$.}
	\label{fig:10}
\end{figure*}

As mentioned above, the non-perturbative proof of relation $\eta = 0$ for the ground
state of the system with Hamiltonian (\ref{2b}) was provided in
\cite{PhysRevB.100.121109,andolina22}. When solving the corresponding Schrödinger equation numerically, we obtain the same result due to the matrix elements (\ref{qrm-matr})  with fixed $M$ values being non-zero only for the field states possessing the same parity. Therefore, in the state vector (\ref{3}) the index $n$ changes with step 2 and, as a consequence, the product of the coefficients
$C^*_{n-1,M} C_{nM} = 0$. At the same time, the matrix elements of the
operator (\ref{1}) do not possess this property and the general value
$\eta \neq 0$ can be obtained. This result is illustrated by Fig.~\ref{fig:4}, which shows that a phase transition occurs for the DM, but is absent in the case of the GIDM. As first demonstrated in the work \cite{PhysRevB.100.121109}, this is due to presence of the diamagnetic term in the GIDM Hamiltonian.

However, the average number of photons in the ground state of the
system within both models is different from zero and determined by the
equation:

\begin{eqnarray}
\label{8}
\alpha =  \bar{n}  =  \braket{\Psi|\hat{a}^\dagger \hat{a}|\Psi} =  \sum_n\sum_M n |C_{nM}|^2.
\end{eqnarray}

Fig.~\ref{fig:5} illustrates the dependence of $\alpha$ on the coupling constant with various atom/qubit numbers for both models. Fig.~\ref{fig:5}a shows that the number of photons in the case of the DM is mainly determined by the coherent state with the classical component of the electromagnetic field $\eta =\braket{\hat{a}}$. However, the electromagnetic field in the case of the GIDM is described by an oscillator with zero average and nonzero occupation number, which corresponds to the squeezed state of the electromagnetic field, as shown in Fig.~\ref{fig:5}b. The properties of these states are shown in Fig.~\ref{fig:6}-\ref{fig:phasespace}.

It is remarkable that the systems possess qualitatively different
statistics of photons in the condensed state. This is illustrated in
Fig.~\ref{fig:6}, where we show the probabilities of the zero-photon $|c_0|^2$ , single-photon $|c_1|^2$, and multi-photon $|c_{mult}|^2 = \sum_{n=2}^{N_{tr}} |c_n|^2$ states:
\begin{eqnarray}
|c_n|^2 = \sum_M |C_{nM}|^2. \nonumber
\end{eqnarray}

These quantities play an important role for the generator of
nonclassical states of the electromagnetic
field\cite{faraon2010generation}.

The resulting condensate consists of photons in a nonclassical
"squeezed" state. Fig.~\ref{fig:7} illustrates the normalized average
squared fluctuations of the coordinate, momentum of the field
oscillators, and the 'squeezing' parameter $r$, calculated as:
\begin{widetext}
  \begin{eqnarray}
\label{9}
  \sigma_x
  =\sqrt{\braket{\hat{x}^2} - \braket{x}^2}, \nonumber \quad  \sigma_p =\sqrt{\braket{p^2} - \braket{p}^2};\nonumber
  \\
  \braket{\hat{x}^2}
  = \frac{1}{2}\braket{(\hat{a} + \hat{a}^\dagger)^2} =\frac{1}{2}\sum_n\sum_M \biggl[\sqrt{n(n-1)}C^*_{n-2,M} + (2n+1)C^*_{n,M}  +\sqrt{(n+1)(n+2)}C^*_{n+2,M}\biggr]C_{nM}; \nonumber
  \\
    \braket{\hat{p}^2}  = - \frac{1}{2} \braket{(\hat{a} - \hat{a}^\dagger)^2}
    =- \frac{1}{2}\sum_n\sum_M
    \biggl[\sqrt{n(n-1)}C^*_{n-2,M} - (2n+1)C^*_{n,M} +\sqrt{(n+1)(n+2)}C^*_{n+2,M}\biggr]C_{nM};
  \nonumber\\
  \alpha -  \eta^2  = \sinh^2 r.
\end{eqnarray}
\end{widetext}

Fig.~\ref{fig:phasespace} illustrates the phase space of the photon condensate for the DM and GIDM. Fig.~\ref{fig:8} shows the population of the photon states at various coupling constants for both models.

Let us stress, that along with the appearance of the photon condensate in the system there
is a transition of a part of atoms to the excited state due to the
interaction with the electromagnetic field in the resonator. The
resulting population of the excited state of atoms can be calculated as:
\begin{eqnarray}
\label{10}
\xi = \braket{\frac{1 + \hat{\sigma}_z}{2}} = \frac{1}{N} \sum_i \braket{\frac{1+\hat{\sigma}_{zi}}{2}}= \nonumber\\\frac{1}{2}+\frac{1}{N} \braket{\hat{J}_z} =
\frac{1}{2} +  \frac{1}{N} \sum_n\sum_M M C^*_{nM} C_{nM}
\end{eqnarray}
\noindent and the corresponding dependence of this value on the coupling constant is shown in Fig.~\ref{fig:9}

Another important characteristic of the system is the entanglement
$\Pi(f,N)$ between the atoms/qubits and the field \cite{PhysRevA.73.062306}:
\begin{eqnarray}
\label{11}
\Pi(f,N)= - \Sigma_M p_M\log_2p_M; \quad p_M = \Sigma_n |C_{Mn}|^2,
\end{eqnarray}
which is illustrated in Fig.~\ref{fig:10}.

\section{Conclusion}
\label{sec:conclusion}

In our work we investigated numerically the possibility of the phase transition within the Dicke model (DM) and gauge-invariant Dicke model (GIDM). This transition was expected to arise from the interaction of an ensemble of two-level systems (atoms/qubits) with the electromagnetic field in a resonator. The obtained results resolve the contradictions about the existence of a phase transition in the DM \cite{PhysRevLett.131.113601} and no-go theorems \cite{PhysRevB.100.121109,andolina22} (and citations
therein). We conclude that for the system consisting of natural two-level objects (for example, particles with spin 1/2 in a magnetic field or artificially created qubits) the phase transition is possible and validity of the DM Hamiltonian for its description is proved. However, the DM becomes invalid when used for modeling the interaction of the system of multi-level objects with the electromagnetic field involving resonant transitions. In such cases, no phase transition occurs and the GIDM should be used instead. Nevertheless, in both cases, the collective
excitation of the field in the system represents a photon
condensate. However, in case of the DM it includes the coherent
quasiclassical states, whereas for the GIDM it consists of nonclassical
"squeezed" states. The computed characteristics of these states are
essential for various applications of QED in the resonators.

\section{Acknowledgements}
AL acknowledges support from DESY (Hamburg, Germany), a member of the Helmholtz Association HGF.

\newpage

\bibliography{biblqrmdm}
\bibliographystyle{abbrv}
\bibliographystyle{unsrt}

\end{document}